\documentclass[%
reprint,
superscriptaddress,
 amsmath,amssymb,
 aps,
prb,
]{revtex4-1}

\usepackage{color}
\usepackage{graphicx}% Include figure files
\usepackage{dcolumn}% Align table columns on decimal point
\usepackage{bm}% bold math
\begin{document}

%\preprint{APS/123-QED}

\title{Magnetic Weyl semimetals with diamond structure realized in spinel compounds}% Force line breaks with \\

\author{Wei Jiang}
\affiliation{Department of Electrical $\&$ Computer Engineering, University of Minnesota, Minneapolis, Minnesota 55455, USA.}%Lines
%\affiliation{Department of Materials Science $\&$ Engineering, University of Utah, Salt Lake City, UT 84112, USA}
%break automatically or can be forced with \\

\author{Huaqing Huang}
\affiliation{Department of Materials Science $\&$ Engineering, University of Utah, Salt Lake City, UT 84112, USA}

\author{Feng Liu}
\affiliation{Department of Materials Science $\&$ Engineering, University of Utah, Salt Lake City, UT 84112, USA}

\author{Jian-Ping Wang}
\affiliation{Department of Electrical $\&$ Computer Engineering, University of Minnesota, Minneapolis, Minnesota 55455, USA.}

\author{Tony Low}
\email[Corresponding author:]{tlow@umn.edu}
\affiliation{Department of Electrical $\&$ Computer Engineering, University of Minnesota, Minneapolis, Minnesota 55455, USA.}%Lines break automatically or can be forced with \\

\date{\today}% It is always \today, today,

\begin{abstract}
Diamond-structure materials have been extensively studied for decades, which form the foundation for most semiconductors and their modern day electronic devices. Here, we discover a $e_g$-orbital ($d_{z^2}, d_{x^2-y^2}$) model within the diamond lattice ($e_g$-diamond model) that hosts novel topological states. Specifically, the $e_g$-diamond model yields a 3D nodal cage (3D-NC), which is characterized by a $d$-$d$ band inversion protected by two types of degenerate states (i.e., $e_g$-orbital and diamond-sublattice degeneracies). We demonstrate materials realization of this model in the well-known spinel compounds (AB$_2$X$_4$), where the tetrahedron-site cations (A) form the diamond sub-lattice. An ideal half metal with one metallic spin channel formed by well-isolated and half-filled $e_g$-diamond bands, accompanied by a large spin gap (4.36 eV) is discovered in one 4-2 spinel compound (VMg$_2$O$_4$), which becomes a magnetic Weyl semimetal when spin-orbit coupling effect is further considered. Our discovery greatly enriches the physics of diamond structure and spinel compounds, opening a door to their application in spintronics.
\end{abstract}

\maketitle

%\tableofcontents
Known as a new classification of quantum matter~\cite{wen2017}, quantum topological ordered systems have attracted tremendous attention for exotic physics (e.g., various quantum Hall effect~\cite{kane2005,hasan2010}, topological superconductivity~\cite{qi2011}, Weyl fermions~\cite{yan2017}) and also for promising applications (e.g., spintronics~\cite{Pesin2012} and quantum computing~\cite{Moore2009}). %It has also shown great potential to bridge the high energy physics such as Weyl and Majorona fermions or even condensed matter analogue of black hole~\cite{fu2008,lutchyn2010,Volovik2016,huang2018}.
In particular, Weyl semimetal materials, characterized with momentum space separated monopole pairs, Fermi arc, and the chiral anomaly, have recently been explored extensively both experimentally and theoretically, as a novel member of the topological family~\cite{burkov2011,lv2015,Shekhar2015}. From the inversion-symmetry breaking TaAs family to the time-reversal symmetry breaking magnetic Heusler alloys and Shandite compounds of $M_{3}M_{2}^{'}X_{2}$, a few families of Weyl semimetals have been theoretically proposed and experimentally confirmed~\cite{wan2011,xu2011,felser2015,wang2016,Liu2018}. However, most of the studied magnetic Weyl semimetals have either both spin channels entangled near the Fermi level or Weyl points formed far away from the Fermi level~\cite{felser2015,wang2016}, which adds difficulties in studying the intrinsic properties of these Weyl points. It is valuable to search for magnetic Weyl semimetal systems as ``clean" as the graphene for Dirac fermions, where the Weyl points is located at the Fermi energy and topological properties can be directly explored and potentially utilized for realistic applications.

%, which add difficulty to understand the underlying physics
The beauty of physics always lie in the simplest models. For example, studies of two-dimensional (2D) graphene have directly spawn various intriguing physical phenomena~\cite{castro2009}, such as Dirac fermion~\cite{Novoselov2005}, quantum Spin Hall effect~\cite{kane2005}, and superconductivity~\cite{uchoa2007,Cao2018}. It has also stimulated many branches of exciting researches, such as silicene, germanene as its 2D analogues~\cite{vogt2012}, valleytronics that utilize $k$ and $k'$ valley degree of freedom~\cite{gunlycke2011,Jiang2013}, and $p_{x,y}$-orbital counterpart of graphene with proposed flat band and Wigner crystallization~\cite{wu2007,Jiang2019}. The diamond structure, which is essentially formed by stacking buckled honeycomb lattice in the A-B-C configuration, can be viewed as a three-dimensional (3D) analogue of the 2D honeycomb lattice, each containing two atoms in one unit cell. It is therefore the simplest system to study analogous exotic physics in three dimensions. Based on diamond structure, theoretical proposal has indeed been made to achieve topological insulators in three dimensions~\cite{fu2007}, whose proposals has yet been able to be mapped onto real materials. %Though have not been explicitly analyzed, recently proposed topological properties of Heusler alloys could potentially be related to their diamond-like structures~\cite{Chadov2010,Lin2010,wang2016}.
It would be important to identify if there exist such an underlying physical model based on diamond structure that leads to various intriguing topological states and can be realized in real material systems.

One of the most famous family of materials that holds the same group symmetry as diamond structure (Fd-3m) is the spinel compounds (AB$_2$O$_4$)~\cite{Mcclure1957,baltzer1965,sickafus1999}. With remarkable magnetic properties, spinel compounds has been extensively studied for decades for their promising appliations in a broad range, e.g, simple permanent magnets, power handling, and magnetic recording et al~\cite{Raul2012}. Surprisingly, its topological properties have been greatly overlooked~\cite{xu2011}. Careful examination shows that there are several advantages studying the topological properties of spinel compounds: 1) ions located at A sites of the crystal structure form a perfect diamond structure; 2) For most of the transition metal ions in A sites, the localized $d$ orbitals are widely seperated from other orbitals, which could yield ideal band structure with clean topological features; 3) The versatility and high tunability of spinel compounds provide a wide range of material candidates with either magnetic or nonmagnetic options; 4) Mature synthesize technology will facilitate its future application in real devices.
%Recently, magnetic Heusler alloys have attracted lots of attention because of their high tunability and variety of topological phases~\cite{Chadov2010,Lin2010,wang2016}. With carefully analysis, they also share the similar diamond structure, especially the half-Heusler alloy (MM$^'$X), where M and M$^'$ form the diamond lattice. Nevertheless, various $d$ orbitals entangled together around the Fermi level, because of the degeneracy and localization nature of the $d$ orbitals, which is very hard to extract a simple model for these topological phases, especially Weyl semimetal phases.

Here, based on tight-binding analysis, we develop a novel two $e_g$-orbitals ($d_{z^2}$ and $d_{x^2-y^2}$) model on the diamond structure ($e_g$-diamond model), as a 3D analogue of the 2D $p_{x,y}$-graphene model~\cite{wu2007}. When both nearest-neighbor (NN) and next-NN (NNN) interaction are considered, the $e_g$-diamond bands form a Dirac/Weyl nodal cage (NC) in the 3D Brillouin zone with hourglass fermions corresponding to the Dirac/Weyl points in the 2D case. The hourglass fermions involved are protected by the coexistence of $e_g$ orbitals degeneracy and diamond crystal symmetry with A, B sublattices degeneracy.
%There exist two types of Dirac/Weyl rings with one type forms in each of the six hexagonal first-Brillouin zone boundaries and the other type forms in each of the $k_{xy,yz,or xz}$ planes crossing the $\Gamma$ point.
Furthermore, we demonstrate a real material realization of the model in a representative 4-2 spinel compounds (V$^{4+}$Mg$^{2+}_2$O$_4$) using first-principles calculations. The spin-polarized $e_g$-diamond bands are exactly half-filled and isolated from other bands because of the strong tetrahedron crystal field splitting and the exchange splitting. The spin-orbit coupling effect further breaks the degeneracy of the 3D-NC, leading to the magnetic Weyl semimetal state. %, as confirmed by topological surface state, Fermi arc and spin texture, and chirality calculations.
It is exciting to note that VMg$_2$O$_4$ has excellent lattice matching (and chemical compatibility) with MgO, which is a widely used spintronics oxide material in industry. This opens the door for the realization of novel spintronics devices, such as achieving low switching energy magnetic devices.
%, We further expand our discovery to a family of spinel compounds. The high tunability of the spinel compounds and their promising integration with the MgO films due to their matching lattice constants and structural similarity make it a promising platform to both study underlying intriguing physical mechanisms and explore potential applications.

%Different from the $p_{x,y}$-graphene model~\cite{wu2007}, the top and bottom flat bands become dispersive in the $e_g$-diamond model.

\begin{figure}
\includegraphics[width=\linewidth]{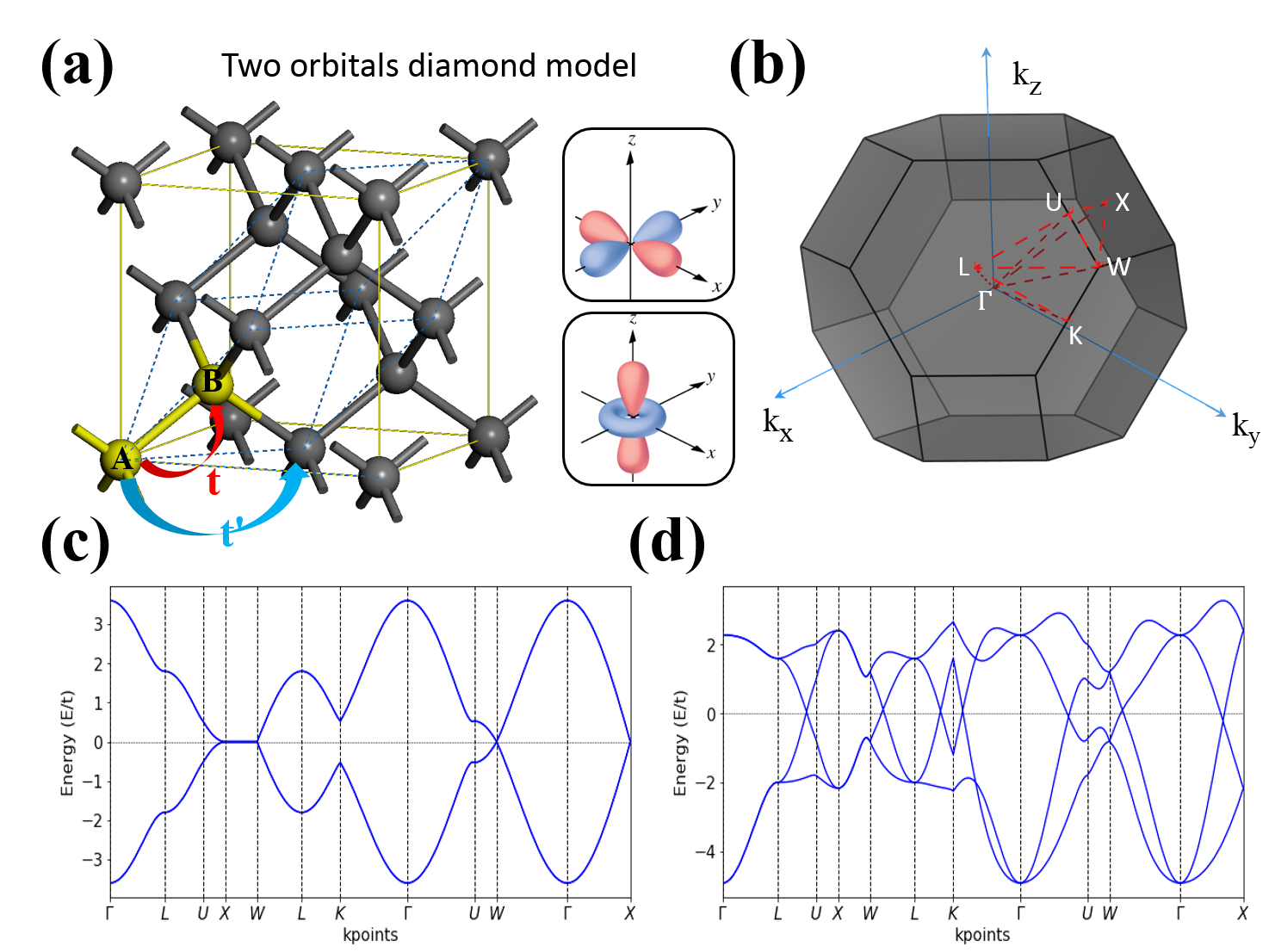} % this command will be ignored
\caption{\textbf{Topological semimetal in $e_g$-orbital diamond model.} \textbf{a} Two-orbital diamond model with nearest-neighbor $t$ and next nearest neighbor hopping $t'$. Insets show the $d_{z^2}$ and $d_{x^2-y^2}$ orbitals for the given coordinates. \textbf{b} First-Brillouin zone with high-symmetry k-points and k-paths. \textbf{c} Band structure of double degenerate single-orbital diamond model with only the NN hopping along high-symmetry paths shown in \textbf{b}. \textbf{d} Band structure of $e_g$-diamond model with nonzero NNN hopping $t'$, showing various linear crossing around the Fermi level.}
\end{figure}

\section*{Tight binding model}

%Sections can only be used in Articles.  Contributions should be organized in the sequence: title, text, methods, references, Supplementary Information line (if any), acknowledgements, interest declaration, corresponding author line, tables, figure legends.

Different from the well-known $sp^3$ or $sp^3s^*$ diamond model, we present here a 3D $e_g$-diamond model. %, analogue to the 2D $p_{x,y}$-graphene model.
For a diamond structure (space group Fd-3m) with the face-centered cubic (FCC) Bravais lattice, there are two equivalent atomic sites in each unit cell (blue dashed line), labelled as A and B, as shown in Fig.1\textbf{a}. By choosing the coordinates in Fig.1\textbf{a} as the orbital quantization axes, we select two energetically degenerate atomic $d_{x^2-y^2}$ and $d_{z^2}$ orbitals ($e_g$ orbitals due to local tetrahedral crystal field splitting) on each site for the $e_g$-diamond model. Without including the spin degree of freedom, this $e_g$-diamond model can be essentially described by a four-band Hamiltonian. It can be viewed as a 3D extension of the 2D $p_{x,y}$-graphene four-band model, in which each honeycomb site is occupied by atomic $p_x$ and $p_y$ orbitals~\cite{wu2007}. To succinctly demonstrate the physics, we limit our Hamiltonian to only the essential NN ($t$) and NNN ($t'$) interactions, and the Hamiltonian can be written as:
\begin{equation}
\mathcal{H} = \sum_i \epsilon_i d_i^\dagger d_i + \sum_{\langle i,j\rangle} t_d d_i^\dagger d_j + \sum_{\langle \langle i,j\rangle\rangle} t^{'}_{ij} d_i^\dagger d_j + H.c. ,
\label{eq:Lieb}
\end{equation}
where $\epsilon_i$ represents the on-site energy of state at $i$ site; $d_i^\dagger$ and $d_i$ are the creation and annihilation operators of $d$ electrons at the site $i$, respectively; $t_d$ and $t_{ij}^{'}$ are the NN and NNN hoppings, respectively. In the diamond structure, each atom has four NNs and twelve NNNs. Given the strict orthogonality between the two $e_g$ orbitals, the NN hopping term between $d_{z^2}$ and $d_{x^2-y^2}$ is set as zero. The non-zero NN hopping terms are among $d_{z^2}$ or $d_{x^2-y^2}$ orbitals, which have the same hopping amplitude $t_d$ for different $e_g$ orbitals and also along different directions due to the geometric isotropic nature of those orbitals along four tetrahedron directions. More extensive TB-model with parameters based on Slater-Koster matrix also confirms these features (Supplementary note 1).

In the absence of the NNN interactions, the same amplitude between hoppings among $d_{z^2}$ orbitals and that among $d_{x^2-y^2}$ orbitals guarantees the double degeneracy of the bands. This can be understood as two exact copies of single-orbital diamond model due to the degeneracy of two $e_g$ orbitals on each site, as can be seen from the band structure shown in Fig.1\textbf{c} (Supplementary note 1). On the other hand, in the single-orbital diamond model, the same hopping amplitude along different directions enables the formation of Dirac nodal lines along X-W path because of the six 2-fold rotation symmetries ($C_2$ axis along x, y, and z directions through A and B sites) in the point group m3m ($O_h$). Therefore, as shown in Fig.1\textbf{c}, the NN $e_g$-diamond model produces a double degenerate two band structure with 4-fold degenerate lines along X-W k-path. It is straightforward that X-W and its inversion and rotational symmetric paths (four 3-fold rotation axis along tetrahedral bond direction) will have the same degeneracy. It is important to emphasize that there exist two types of band degeneracies: one type is due to the degeneracy of $e_g$ orbitals (named type-A thereafter, $d_{z^2}$ and $d_{x^2-y^2}$) and the other type owing to the $C_2$ rotational symmetry in the single-orbital diamond Hamiltonian (named type-B thereafter, $d_{z^2}$ or $d_{x^2-y^2}$ in A and B sites).  %\textcolor{red}{to test the other C$_4$ rotational path X W' all the six rotational symmetry have to be conserved at the same time}

Next, we include the NNN hopping interactions, where NNNs form the FCC structure. There are two types of NNN hoppings, i.e., hoppings between $d_{z^2}$ and $d_{x^2-y^2}$ orbitals (termed type-I thereafter, $t_1$), and hoppings among $d_{z^2}$/$d_{x^2-y^2}$ orbitals (termed type-II thereafter, $t_2$). Considering each single-orbital diamond Hamiltonian as one block, the type-I hopping terms mix the two unit blocks by introducing the off-diagonal interaction, and the type-II hopping terms act as on-site energy variation of the two blocks (Supplementary note 1). Let us examine effects of these two types of NNN interactions to the band structure. First, it is apparent that if $t_2$ remains the same for $e_g$ orbitals, i.e., same variation of the on-site energies, the band degeneracy remains, leading to a similar band structure with slightly different dispersion (Supplementary Fig. 2). Only different $t_2$ can lift the type-A degeneracy of those two bands. On the other hand, any non-zero $t_1$ will lift the type-A degeneracy due to the off-diagonal orbital mixing.

Interestingly, we notice that under some symmetric hopping conditions, the band degeneracy along certain high-symmetry k-paths remains even for nonzero $t_1$ and different $t_2$ NNN hoppings. Specifically, we find that at the center point of the eight hexagonal Brillouin boundary ($K_{HC}$), the type-A degeneracy remains when the NNN hopping Hamiltonian keeps the $C_4$ rotational symmetry. Similarly, at the $\Gamma$ and any point between $\Gamma$ and $K_{HC}$, the band degeneracy will remain while both $C_4$ rotational symmetry and the summation of hoppings along 12 NNN (3 nonequivalent) directions equals to zero (Supplementary note 1 and Supplementary Fig. 3). Note the type-B degeneracy along X-W path is also conserved as the $C_2$ rotation symmetry remains. One important outcome of the coexistence of two types of band degeneracy is the guaranteed band crossing (nodal point, hourglass fermion) along certain k-paths, when the two terminal k-points have different types of degeneracies, as shown in Fig.~1\textbf{b}. Such band crossing can be understood from the continuity nature of the wavefunctions that essentially leads to a $d$-$d$ band inversion, since the two degenerate bands are formed with different orbital states (Supplementary Fig. 3). As a consequence, those nodal points surprisingly construct a three-dimensional nodal cage, which will be further discussed later.

\section*{Magnetic semimetal in spinel compounds}

To realize the aforementioned $e_g$-diamond model in real materials, we focus our attention on the well known spinel compounds that also have a space group of Fd-3m as that of the diamond structure~\cite{Mcclure1957,baltzer1965,sickafus1999}. These inorganic oxides, with a chemical formula of AB$_2$O$_4$, are constructed by FCC lattice of $O^{2-}$ anions and interstitial cations in tetrahedral (T-site) and octahedral (O-site) sites formed by $O^{2-}$ ions, as shown in Fig.2\textbf{a}. Sulphur or other chalcogenides could also be used as anions, and the corresponding compounds are normally termed as thiospinel. There are eight tetrahedral and four octahedral sites in the spinel compounds. In a normal spinel structure, cations A occupy one-eighth of the T-sites and cations B occupy one-half of the O-sites. There are also inverse spinel and mixed spinel structures that have A and B ions filling different O- and T- sites. Here, we will only focus on the normal spinel compounds with each types of polyhedron site occupied by the same type of metal cations. These normal spinels can be further classified into two categories based on the valence states of the cations, i.e., 2-3 spinel with A$^{2+}$ and B$^{3+}$; 4-2 spinel with A$^{4+}$ and B$^{2+}$. It is important to mention that the A cations in T-sites form exactly a diamond structure and B cations in O-sites form a 3D kagome lattice instead. The 3D kagome lattice has been well-documented in other oxide compounds, such as pyrochlore (Fd-3m) with the chemical formula of A$_2$B$_2$O$_7$~\cite{gardner2010}, which was predicted to hold magnetic Weyl semimetal state and 3D flat band~\cite{wan2011,Witczak-Krempa2012,Zhou2019}. However, to the best of our knowledge, though the diamond structure made by transition metal ions exist in many compounds, their intriguing electronic and topological properties have not been studied.

\begin{figure}
\includegraphics[width=\linewidth]{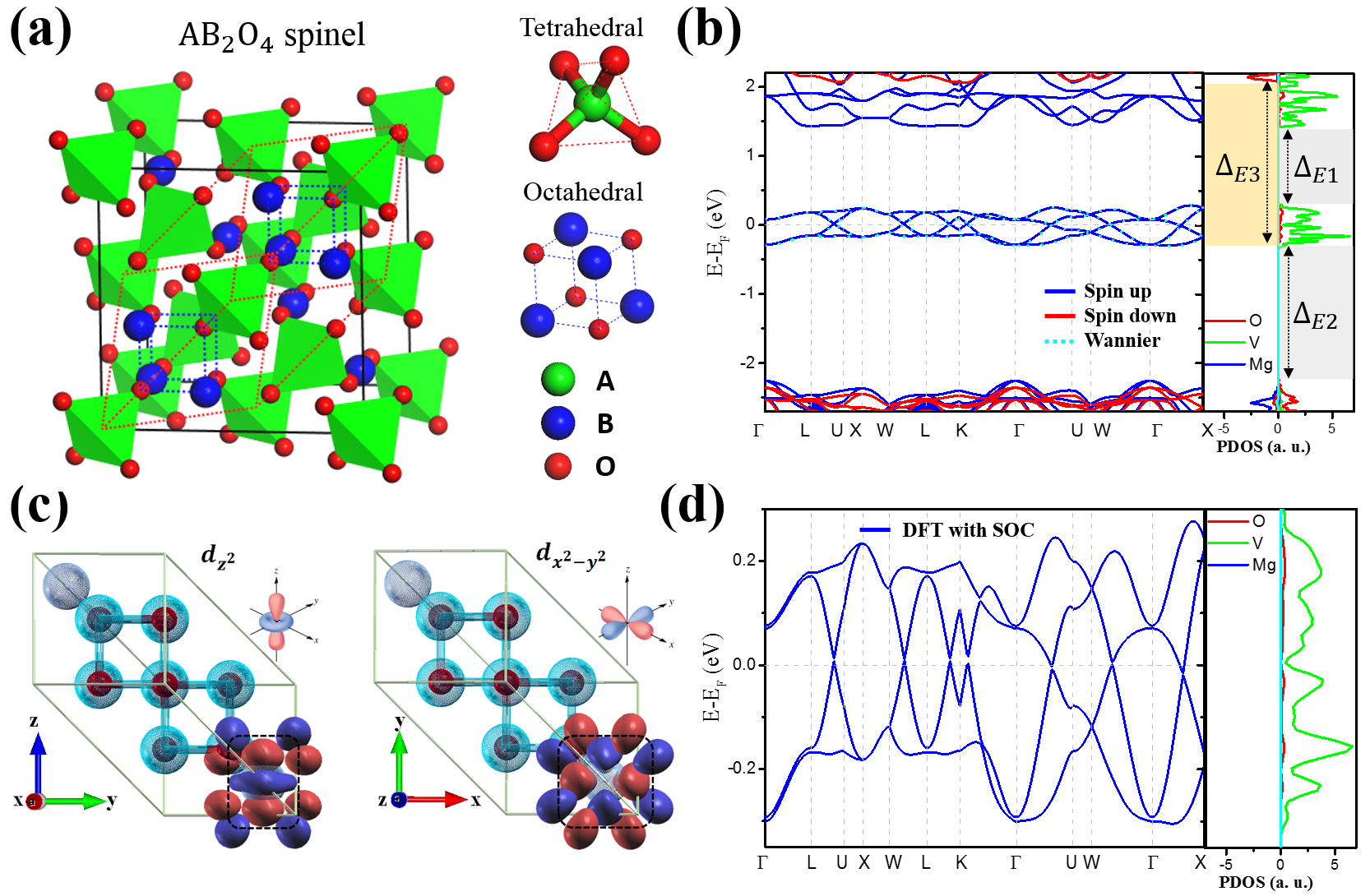} % this command will be ignored
\caption{\textbf{Electronic and magnetic properties of VMg$_2$O$_4$.} \textbf{a} Conventional atomic structure of an ionic spinel compound with chemical formula of AB$_2$O$_4$ with the primitive unit cell labelled using dashed red lines. Cations A and B occupy the tetrahedral and octahedral sites, respectively. \textbf{b} Band structure of 4-2 spinel (V$^{4+}$Mg$_2^{2+}$O$_4$), showing half-filled four-band structure  isolated from the other bands with $d-d$ crystal splitting gap $\Delta_{E1}$, $p-d$ charge transfer gap $\Delta_{E2}$, and exchange splitting gap $\Delta_{E3}$. \textbf{c} Maximally localized wannier functions of $d_{z^2}$ and $d_{x^2-y^2}$ orbitals from wannier fitting. \textbf{d} Enlarged four-band structure near the Fermi level from DFT and the wannier fitting, respectively. }
\end{figure}

For A cations, the local tetrahedral crystal field splits the five degenerate $d$ orbitals into three-fold degenerate $t_{2g}$ and two-fold degenerate $e_g$ orbitals with relatively lower energies. Further exchange splitting of these $d$-orbitals will break the time reversal symmetry and lift the spin degeneracy (Supplementary note 2 and Supplementary Fig. 4). With proper filling of the $e_g$ orbitals, e.g., one electron per site, it could form a half-filled magnetic $e_g$-diamond bands around the Fermi level. Indeed, we found one experimentally studied 4-2 spinel compound with chemical formula of VMg$_2$O$_4$ that perfectly satisfies these criteria~\cite{oshima1977,hellmann1983}, where vanadium cations (4$s^2$3$d^3$) have one $d$ electron left after donating four valence electrons (V$^{4+}$) to neighboring oxygen anions (O$^{2-}$). Therefore, we carried out first-principles calculation (see methods) to study its electronic and magnetic properties. The calculations show that VMg$_2$O$_4$ is a ferromagnetic (FM) material with a magnetic moment of 2 $\mu_B$ per unit cell (UC), which are mainly contributed by the V cations, as confirmed by the spin distribution plot (Supplementary Fig. 5). The energy difference between FM and antiferromagnetic (AFM) state is around 0.43 eV at the PBE level, indicating a promising room temperature FM feature (Supplementary note 3 and Supplementary Fig. 5).

We first calculated the band structure of VMg$_2$O$_4$ without considering the spin-orbit coupling (SOC) effect. As shown in Fig.2\textbf{b}, there is a clear four-band structure right at the Fermi level with very good isolation from other bands. The band structure exhibits an ideal half-metallicity feature where one spin channel is metallic and the other spin channel is insulating with a band gap as large as 4.36 eV at the PBE level of accuracy. We further calculated the system using hybrid functionals, which shows the same feature with a even larger spin gap (5.17 eV for meta-GGA and 6.62 eV for HSE, Supplementary Fig. 6). As we are interested in the four-band structure, we will use the PBE results for simplicity. To understand this peculiar electronic structure, we plotted and analyzed the projected density of states (PDOS) of VMg$_2$O$_4$. As shown in Fig.2\textbf{b}, the four bands around the Fermi level are nearly half-filled and mainly contributed by V $d$ electrons, which is consistent with the magnetic moment distribution. The PDOS of Mg ions are found located far above the Fermi level, as the two valence electrons are fully transferred to O$^{2-}$ ions to form the Mg$^{2+}$ with zero valence electrons. The oxygen state are about 2 eV below the Fermi level because of the closed shell electronic configuration after gaining electron from Mg and V. % These features are further qualitatively confirmed from the Bader charge analysis (Supplementary Table 1).

The observed band isolation in spin up channel can be traced to the underlying energy splittings indicated as $\Delta_{E1}$ and $\Delta_{E2}$, as shown in Fig.2\textbf{b}. The energy gap above ($\Delta_{E1}$) is between $d$ orbitals, which essentially is the tetrahedral crystal field splitting energy ($\Delta_{tet}$) between $e_g$ and $t_{2g}$ orbitals. The large separation, $\Delta_{tet}$ $\approx$ 2 eV, is due to the strong interaction with $O^{2-}$ ions and the high oxidation state of V$^{4+}$. The energy gap below the Fermi level ($\Delta_{E2}$) is between $d$ and $p$ orbitals, which is the charge-transfer gap determined basically by the energy-level separation of orbitals between vanadium and  oxygen ligand. The separation can be well characterized by the ligand electronegativity and the Madelung potential~\cite{Ohta1991}. On the other hand, the large band gap in the spin down channel can be understood as the cooperative effect of $\Delta_{E2}$ and large $\Delta_{E3}$. $\Delta_{E3}$ is essentially the exchange splitting of the V $d$ electrons, which is known to be significant because of the strong localization effect of transition metal $d$ orbitals. Ideally, such large gap would guarantee pure spin current and prevent any current leakage of the minority spin. To the best of our knowledge, this is also the largest spin gap among reported half-metallic ionic compounds.

\section*{$E_g$-diamond model and 3D-NC in VMg$_2$O$_4$}
More interestingly, a closer look at these four bands (Fig.2\textbf{b}) reveals multiple linear crossings around the Fermi level, indicating a semimetallic feature. This is also supported by the enlarged PDOS plot, which shows nearly zero DOS around the Fermi level (Fig.2\textbf{d}). Besides those crossings (nodal points), there also exist several degenerate bands along $\Gamma$-L, X-W. %and their inversion and rotational symmetry paths.
These features agree perfectly with our $e_g$-diamond model, and the slight difference of band dispersion is simply due to the different hopping integrals, which will not affect the crossing features of interest. To confirm the $e_g$-diamond model, we performed the maximally localized Wannier functions (MLWFs) calculation to fit to the DFT band structures using the Wannier90 package~\cite{Marzari1997}. The MLWFs fitted band structure based on two $e_g$-orbitals on diamond lattice agrees perfectly with the DFT result, as can be seen in Fig.2\textbf{b}. Furthermore, the calculated MLWFs show clearly the shape of $d_{z^2}$ and $d_{x^2-y^2}$, as shown in Fig.2\textbf{c}, confirming the orbital characteristics and again validating the $e_g$-diamond model. This is further supported from the band-resolved partial charge distribution plot of these four bands, which show a characteristic shape composed of the two orbitals (Supplementary Fig. 7).

As demonstrated from the $e_g$-diamond model, we expect the band structure of VMg$_2$O$_4$ will also form a nodal cage in the 3D first-Brillouin zone. To confirm the 3D nodal cage feature, we first calculated 3D band structures in two 2D k-planes with one plane at the hexagonal Brillouin boundary [(111) surface] and the other one at k$_x$-k$_y$ plane across the $\Gamma$ point [(110) surface], as shown in Fig.3\textbf{a} and \textbf{c}, respectively. The 3D band structure (Supplementary Fig. 8) and the band gap between the middle two Dirac bands (Fig.3\textbf{b} and \textbf{d}) show clearly nodal ring features within both the hexagon (111) and octagon (110) plane, respectively. According to the C$_4$ rotational symmetry and the inversion symmetry, the other seven hexagonal Brillouin zone boundaries also possess the same feature as (111) surface (Supplementary Fig. 8). Similarly, symmetry invariant planes, i.e., (101) and (011) planes across the $\Gamma$ point, of (110) surface possess the same feature due to the C$_3$ rotational symmetry (Supplementary Fig. 8).

From the band structure in Fig.2\textbf{b}, we can see nearly all the nodal points are located closely adjacent to the Fermi level ($<$ 20 meV). Therefore, the Fermi surface could essentially capture the 3D nodal cage feature. Indeed, the Fermi surface plot, as shown in Fig.3\textbf{e}, shows a smooth 3D surface within the 3D Brillouin zone that ends at the hexagonal Brillouin zone boundaries, which interestingly has a similar shape as seen in Cu metal~\cite{Segall1962}. Because of the small energy dispersion of the nodal cage, we see the coexistence of electron and hole pockets, where hole pockets are mostly located near the square zone boundaries. The cross sections of the 3D Fermi/nodal cage for different planes (2D Fermi surface, Supplementary Fig. 9 and 10) are also consistent with previously demonstrated nodal rings feature based on 3D band structure calculations. It is found that the nodal line within (111) surface is nearly flat, while that of (110) surface shows clear energy oscillation ($\pm$ 20 meV) as it cut across the square zone boundaries. We also calculated the 3D band structure and 2D Fermi surface for two series of 2D k-planes (Supplementary Fig. 9-11), i.e., (111) planes and (110) planes from the boundary to the center of the Brillouin zone with a step of 0.1 \AA, which further confirm the intriguing 3D nodal cage feature.

\begin{figure}
\includegraphics[width=\linewidth]{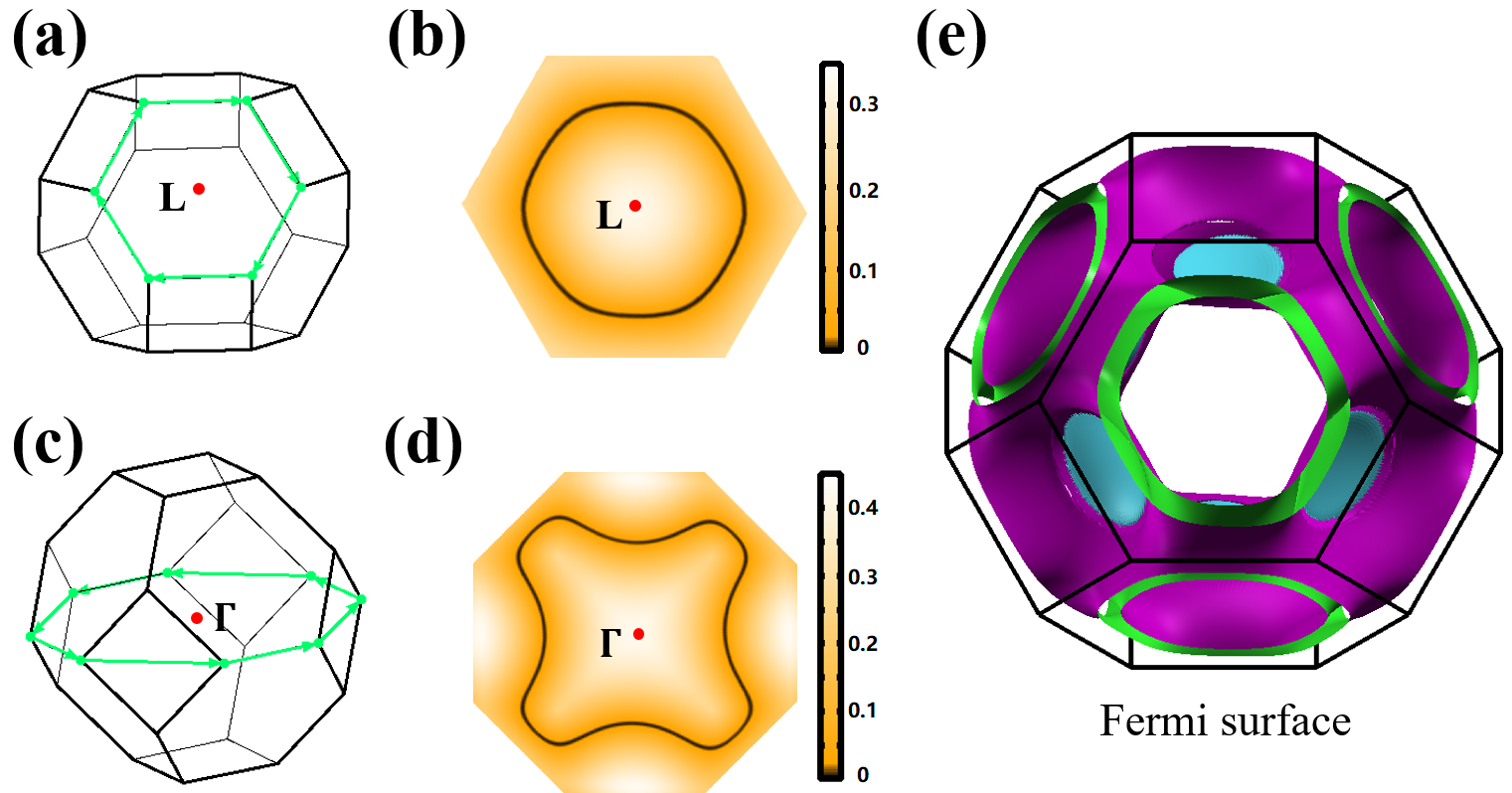} % this command will be ignored
\caption{\textbf{3D Nodal cage.} \textbf{a} 2D hexagonal k-plane of the (111) surface at the Brillouin zone edge. \textbf{b} Band gap between the middle two bands for the 2D k-plane in \textbf{a}. \textbf{c} and \textbf{d} same as \textbf{a} and \textbf{b} for the 2D octagonal k-plane of the (110) surface across the $\Gamma$ point of the 3D Brillouin zone. \textbf{e} Fermi surface at the energy of Fermi level, showing the feature of 3D nodal cage. Purple and blue regions represent the hole and electron pockets, respectively. Green region highlights the cross section at the Brillouin zone boundary.}
\end{figure}

\section*{Magnetic Weyl semimetal}

Generally, such nodal cage is not robust, which will become gapped after considering SOC effect~\cite{huang2018,chen2018}. Depending on whether the system becomes fully gaped or not, the material will become the quantum anomalous Hall insulator or magnetic Weyl semimetal. Therefore, we proceed to include the SOC to determine the topological phases of the spinel compound VMg$_2$O$_4$. Firstly, we performed the non-collinear calculations with different spin configurations (Supplementary note 4), which show that the system continues to be FM with nearly degenerate energies along different magnetization directions. The corresponding band structure for the magnetization direction along $<$111$>$ direction is plotted along high-symmetry k-paths (Fig.2\textbf{d}), where most of the nodal points become gapped with several k-points remaining degenerate, such as that along $\Gamma$-W that is perpendicular to the magnetic field. This is a known effect in magnetic Weyl semimetal system, where the positions of Weyl points are dependent on the magnetization direction~\cite{Zhou2019}. Further detailed calculations show that there exist 18 pairs of degenerate points within the first Brillouin zone, which are confirmed to be Weyl pairs with Berry phase of $\pm \pi$ through the Berry curvature integration around these points (Supplementary Fig. 12 and Supplementary Table 1).

To demonstrate the topological properties of this magnetic Weyl semimetal, we first calculated the surface state using a semi-infinite system based on Green's function. As can be seen from Fig.4\textbf{a} and \textbf{d}, topological non-trivial surface states connecting the bulk states can be clearly seen for both (111) and (110) surfaces, respectively. As one of the characteristic feature of the Weyl semimetals, the Fermi arc for these two surfaces are also calculated, as shown in Fig.4\textbf{b} and \textbf{e}. The Fermi arcs show dramatic difference between different surfaces due to different projection of Weyl pairs. For the (110) surface, two pairs of Weyl points can be clearly seen from the two seperated Fermi arcs as required by the C$_2$ rotational symmetry. For the (111) surface, the Fermi arcs form a ``Fermi ring" with three pairs of Weyl points due to the C$_3$ rotational symmetry. These Weyl points are accidentally overlapping while projecting to the (111) surface, which are essentially separated within the bulk Brillouin zone. %We also calculated the evolution of the Fermi arc at different energy levels, which further confirm the topological feature of the magnetic Weyl semimetal system.
The spin textures of the Fermi arc shown in Fig.4\textbf{c} and \textbf{f} indicate the positive and negative chirality of Weyl points, which are further confirmed from the Berry phase calculation by integrating the Berry curvature along the surface enclosing the Weyl points.

\begin{figure}
\includegraphics[width=\linewidth]{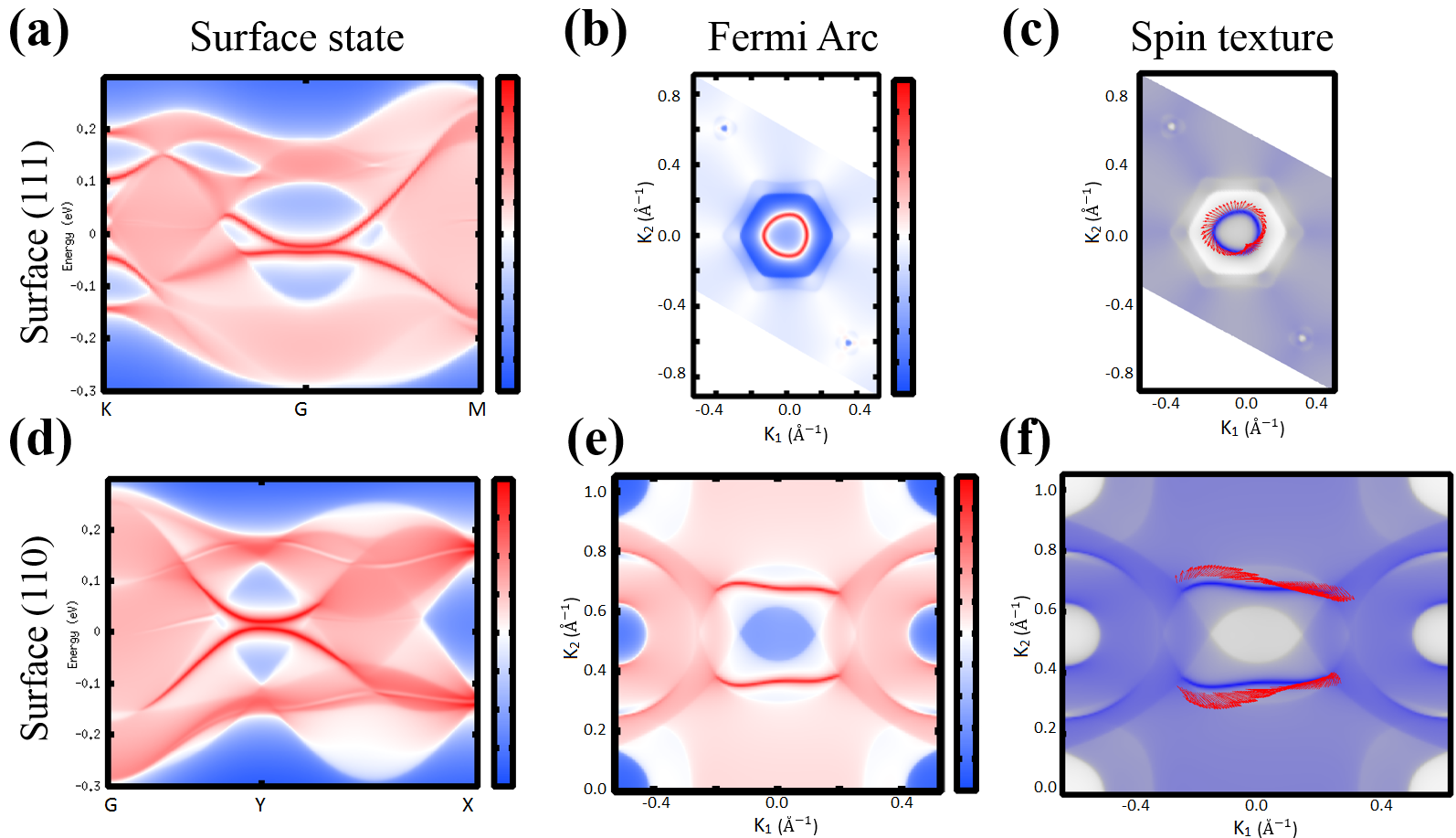} % this command will be ignored
\caption{\textbf{Topological properties of Weyl semimetal VMg$_2$O$_4$.} \textbf{a} Topological surface state connecting bulk states for the (111) surface. \textbf{b} and \textbf{c} Fermi arc and the corresponding spin texture of the (111) surface. \textbf{d}, \textbf{e}, and \textbf{f} Same as \textbf{a}, \textbf{b}, and \textbf{c} for (110) surface.}
\end{figure}

\section*{Discussion and perspectives}

One of the intriguing properties of Weyl semimetal is their proposed large intrinsic anomalous Hall conductivity $\sigma_H^A$, which comes from the integration of Berry curvature of bulk band structure with Weyl points~\cite{Zeng2006,Onoda2006}. Considering the small charge current $\sigma$ due to the semi-metallic feature, it is believed that Weyl semimetal could yield a large anomalous Hall angle $\sigma_H^A/\sigma$. Therefore, we also calculated intrinsic anomalous Hall conductivity of VMg$_2$O$_4$, which shows a relatively large peak ($\approx$ 100 $\Omega^{-1}\cdot cm^{-1}$) around the Fermi level (Supplementary Fig. 13). We note that $\sigma_H^A$ of VMg$_2$O$_4$ is much smaller than the largest $\sigma_H^A$ reported in literature~\cite{Liu2018,Wang2018}. This is reasonable as in VMg$_2$O$_4$, only those nodal points contribute to the $\sigma_H^A$ around the Fermi level. Though the Weyl points may serve as hot spot with extremely large Berry curvature, the contribution from positive and negative Weyl points may cancel each other and lead to a relatively small total $\sigma_H^A$. Theoretical proposals suggest engineering extra bulk bands to cross the Fermi level, which may help contributing to large $\sigma_H^A$~\cite{Derunova2019}, though the system is no longer a Weyl semimetal. To test this idea, we modified the band dispersion through strain engineering and successfully realize the proposed band structure upon compressive strain. We further calculate the $\sigma_H^A$, which indeed show a large enhancement of the $\sigma_H^A$ by at least 2-folds (Supplementary Fig. 13).

The intriguing phenomenon observed in VMg$_2$O$_4$ can also be extended to other spinel compounds. Through theoretical calculations, we find the spinel compounds formed with elements having the same valance states as VMg$_2$O$_4$, e.g., VMg$_2$Se$_4$ and VCa$_2$O$_4$, share the same features as VMg$_2$O$_4$, i.e., magnetic Weyl semimetal phase with $e_g$-diamond model (Supplementary Fig. 14). On the other hand, based on the same model, with different electron filling, some other potentially interesting phases can be realized with elements having different valence states. For example, we can acquire large gap ($\approx$ 3 eV) half-metal phase in MnMg$_2$O$_4$ and realize an FM insulating state in CrMg$_2$O$_4$ (Supplementary Fig. 14). It is interesting to mention that FM insulator possesses two different band gaps (0.6 eV and 3.6 eV) in different spin channels, which could potentially be used as spin filter in magnetic tunneling junction devices.

On the other hand, the lattice compatibility/match with (001) textured MgO tunneling barrier is critical for practical applications. Many materials, such as heusler materials, that have been proposed and/or studied faces practical integration issues with MgO because of lattice mismatch. Interestingly, the material we studied here has a very small lattice mismatch with MgO for both (001) and (111) planes ($<$ 0.4$\%$), because of their structural similarity (Supplementary Fig. 15). This lattice matching might facilitate high quality growth of VMg$_2$O$_4$ directly on top of MgO to study or utilize its intriguing magnetic Weyl semimetal features, as good quality MgO thin films have been easily grown and widely utilized for spintronic devices. For example, we can build MTJ devices with VMg$_2$O$_4$/MgO/VMg$_2$O$_4$ stacking or simply use VMg$_2$O$_4$ as the spin filtering layer (Supplementary Fig. 16). We note that there also exist a well-studied 2-3 vanadium spinel, MgV$_2$O$_4$, which contains the same types of elements but different stoichiometric ratio, valence states, as well as occupation of cations~\cite{Tchernyshyov2004,Lee2004}. MgV$_2$O$_4$ is found to be a spin-1 AFM insulator with strong frustration due to the 3D kagome lattice formed by the magnetic V ions, which is dramatically different from that of VMg$_2$O$_4$ discussed here. In order to get the VMg$_2$O$_4$ phase with high valence state V$^{4+}$ experimentally, relatively high oxygen pressure and Mg:O ratio might be necessary~\cite{oshima1977,hellmann1983}.

\section*{Conclusion}
In summary, we have discovered a novel $e_g$-diamond four-band model which yields intriguing nodal cage feature due to the coexistence of orbital and sublattice degeneracies. We discovered that such model can be realized in the well-studied spinel oxide compounds. Using 4-2 spinel compound VMg$_2$O$_4$ as a representative example, we confirmed the validity of the $e_g$-diamond model and demonstrate the formation of 3D nodal cage due to linear crossing between the middle two bands. Its topological properties are studied, indicating the material to be a magnetic Weyl semimetal, which is novel to the spinel compounds. We further expand the model to a series of spinel compounds and demonstrate their promising applications as spintronic materials. This theoretical discovery substantially enriches the physics of spinel compounds and could potentially lead to new applications.%, which will attract immediate experiment attention.

\textit{Acknowledgement.} This project is supported by SMART, one of seven centers of nCORE, a Semiconductor Research Corporation program, sponsored by National Institute of Standards and Technology (NIST). We thank the MSI at the University of Minnesota for providing the computing resources.

\bibliography{sample}% Produces the bibliography via BibTeX.

\end{document}